\documentclass[a4paper,prd,superscriptaddress,11pt,noshowkeys,notitlepage,nofootinbib]{revtex4}

\usepackage[english]{babel}
\usepackage{mathtools}
\usepackage[]{units}
\usepackage[top=2cm,bottom=2cm,left=3cm,right=3cm,marginparwidth=1.75cm]{geometry}
\usepackage[usenames]{xcolor}
\usepackage{amsfonts,amsmath,amssymb}

\usepackage{ulem}
\usepackage{graphicx}
\usepackage[colorlinks=true, allcolors=blue]{hyperref}
\usepackage{subcaption}
\newcommand{\UPTC}{Escuela de Física, Universidad Pedagógica y Tecnológica de Colombia,\\
Avenida Central del Norte \# 39-115, Tunja, Colombia}
\newcommand{\UdeA}{Instituto de Física, Universidad de Antioquia,\\Calle 70 \# 52-21, Apartado Aéreo 1226, Medellín, Colombia}

\begin{document}

\title{Singlet Dirac dark matter streamlined}

\author{Carlos E. Yaguna}
\affiliation{\UPTC}
\author{\'Oscar Zapata}
\affiliation{\UdeA}
\begin{abstract}
We propose a new and compact realization of singlet Dirac dark matter within the WIMP framework. Our model replaces the standard $Z_2$ stabilizing symmetry with a $Z_6$, and uses spontaneous symmetry breaking to generate the dark matter mass, resulting in a much simplified scenario for Dirac dark matter. Concretely, we extend the Standard Model (SM) with just two new particles, a Dirac fermion (the dark matter) and a real scalar, both charged under the $Z_6$ symmetry. After acquiring a vacuum expectation value, the scalar gives mass to the dark matter and mixes with the Higgs boson, providing the link between the dark sector and the SM particles. With only four free parameters, this new model is extremely simple and predictive. We study the dark matter density as a function of the model's free parameters and use a likelihood approach to determine its viable parameter space. Our results demonstrate that the dark matter mass can be as large as $6$ TeV while remaining consistent with all known theoretical and experimental bounds. In addition, a large fraction of viable models turns out to lie within the sensitivity of future direct detection experiments, furnishing a promising way to test this appealing scenario.
\end{abstract}

\maketitle

\section{Introduction}
Weakly interacting massive particles (WIMPs) still are one of the best motivated candidates to explain the dark matter observed throughout the Universe~\cite{Roszkowski:2017nbc,Arcadi:2017kky}. They can be easily incorporated into many Standard Model (SM) extensions, naturally give rise to a relic abundance of the right order of magnitude~\cite{Planck:2018vyg}, and can usually be probed in current and future dark matter~\cite{Schumann:2019eaa} or collider experiments~\cite{Buchmueller:2017qhf}. Within this WIMP framework, many different dark matter models can be envisaged, depending on the number and type of new fields that are included and on the additional symmetries that may be imposed. 

In this paper we propose a new and simplified scenario for Dirac dark matter. In it, the SM is extended with just two additional fields --a Dirac fermion (the dark matter) and a real scalar, both singlets under the SM gauge group. The main novelty of our model is the use of a  $Z_6$ discrete symmetry rather than the standard $Z_2$ as a stabilizing symmetry for the dark matter particle. Under it, only the new fields transform non-trivially. The resulting scenario is very predictive as it admits only four free parameters, significantly fewer than similar models for Dirac dark matter. After introducing the model and describing its dark matter phenomenology, we use a likelihood approach  to identify its viable parameter space --the regions consistent with all known theoretical and experimental bounds. Our results demonstrate that this scenario is viable over a wide range of dark matter masses extending from about $60$ GeV to $6$ TeV. Moreover, the new scalar turns out to play an essential role: it generates the dark matter mass via its vacuum expectation value; it allows the dark matter to annihilate efficiently in the early Universe so as to satisfy the relic density constraint; and it links the dark matter with the SM fields, providing a way to probe this scenario in current and planned experiments.  This simple setup, therefore, provides a minimal, consistent and testable realization of singlet Dirac dark matter. 

Numerous studies have investigated singlet fermion dark matter models, primarily employing a $Z_2$ symmetry and introducing additional free parameters (e.g.,~\cite{Kim:2008pp,Baek:2011aa,Baek:2012uj,Esch:2013rta,Freitas:2015hsa}). These scenarios have also been analyzed from an effective field theory standpoint (e.g.,~\cite{Kim:2006af,Fedderke:2014wda,Matsumoto:2014rxa,GAMBIT:2018eea}), and the prospect of electroweak baryogenesis has gained recent attention (e.g.,~\cite{Fairbairn:2013uta,Li:2014wia,Ghorbani:2017jls,Beniwal:2018hyi,Gould:2023jbz}). Over the past few years, $Z_N$ symmetries have been integrated into diverse dark matter models, often involving multi-component dark matter (e.g.,~\cite{Yaguna:2019cvp,Belanger:2020hyh,Belanger:2021lwd,Yaguna:2021vhb,Yaguna:2021rds,Choi:2021yps,DiazSaez:2022nhp,Jurciukonis:2022oru,Belanger:2022esk,Dupuis:2016fda,Bhattacharya:2018ljs,Ho:2022erb,Yaguna:2023kyu}). WIMP models using a $Z_4$ with a singlet scalar mediator featuring a Dirac and a Majorana dark matter candidate were analyzed in Ref.~\cite{Dupuis:2016fda} and Ref.~\cite{Bhattacharya:2018ljs}, respectively (in Refs.~\cite{Ho:2022erb} and~\cite{Yaguna:2023kyu} alternative dark matter production mechanisms were implemented). 
To our knowledge, the $Z_6$ model for Dirac dark matter we focus on is novel. 

The rest of the paper is organized as follows. In the next section the model is introduced and its free parameters are identified.  The dark matter phenomenology is discussed in section \ref{sec:pheno}, where the variation of the relic density with the model's parameters is analyzed. Section \ref{sec:pspace} presents our main results. There, the viable regions of this model are determined and analyzed, and the detections prospects are investigated. In section \ref{sec:disc} we contrast this model against related scenarios considered previously in the literature, and some of its possible extensions are mentioned. Finally, our conclusions are presented in section \ref{sec:conc} . 
\section{The Model}
\label{sec:mod}
To account for the dark matter, we propose to extend the SM particle content with just two chiral fermions ($\psi_L$ and $\psi_R$) and a real scalar field ($\phi$), all singlets under the SM gauge symmetry.  In addition, a $Z_6$ discrete symmetry is assumed under which only the new fields transform non-trivially and as follows\footnote{Notice that both fermions must be charged under the $Z_6$ to avoid a Yukawa interaction term with the lepton doublet, e.g., $\bar{L}\tilde{H}\psi_R$.}
\begin{align}
\phi \rightarrow \omega_6^3 \,\phi,\,\,\,    \psi_L \rightarrow \omega_6 \,\psi_L,\,\,\, \psi_R \rightarrow \omega_6^4 \,\psi_R; \hspace{1cm}\omega_6=e^{\frac{2\pi i}{6}}. 
\label{eq:charge}
\end{align}
This particle content and  symmetry transformations specify  our setup. 

A consequence of the above transformations is that bare mass terms for the fermions are forbidden by the $Z_6$ symmetry, either of Dirac or of Majorana type. It is possible, however, to write the following  interaction term between $\psi_L$, $\psi_R$ and $\phi$:  
\begin{align}\label{eq:L}
 \mathcal{L}_\psi&=\,\,-y_s\overline{\psi}_L\psi_R\, \phi+ \rm{h.c.},
 \end{align}
 which will induce a Dirac mass term for the fermions once $\phi$ acquires a vacuum expectation value.   

The tree-level scalar potential for this model is  
\begin{align}\label{eq:V}
 \mathcal{V}_0&=\,\,-\mu_H^2|H|^2+\lambda_H|H|^4-\frac{1}{2}\mu_{\phi}^2\phi^2 + \frac{1}{4}\lambda_{\phi}\phi^4 +\frac{1}{2}\lambda_{S H}|H|^2\phi^2, 
 \end{align}
where $H$ is the SM Higgs doublet. Notice that terms such as $\phi^3$ or $\phi |H|^2$ are forbidden by the $Z_6$ symmetry because $\phi$ transforms to $-\phi$ under it. As explained above,  $\phi$ must acquire a vacuum expectacion value so we write
\begin{align}
H&=\frac{1}{\sqrt{2}}\left(\begin{array}{c}
       0 \\
        H^0_R+v_H 
    \end{array}\right),\hspace{1cm}
    \phi=\left(\phi_R + v_\phi\right). 
\end{align}
After the $Z_6$ and electroweak symmetry breaking the real components $\phi_R$ and $H^0_R$ get mixed through the $\lambda_{SH}$ term. Specifically, the $2\times2$ mass matrix in the  $(H^0_R,\phi_R)$ basis has the entries 
\begin{align}
    M^2_{11}&=2\lambda_H v_H^2,\\
    M^2_{22}&=2\lambda_\phi v_\phi^2,\\
    M^2_{12}&=M^2_{21}= \lambda_{SH}v_\phi v_H.
\end{align}
The mass eigenstates $h,S$ are defined through the rotation matrix
\begin{align}
\begin{pmatrix}
    H^0_R \\
    \phi_R
\end{pmatrix}&=
\begin{pmatrix}
    \cos\theta & \sin\theta \\
    -\sin\theta & \cos\theta
\end{pmatrix}
\begin{pmatrix}
    h \\
    S
\end{pmatrix},
\end{align}
with a mixing angle 
\begin{align}
    \tan 2\theta&=\frac{\lambda_{SH}v_\phi v_H}{\lambda_\phi v_\phi^2-\lambda_Hv_H^2},\\
     \sin 2\theta&=\frac{2\lambda_{SH}v_\phi v_H}{M_S^2-M_h^2},\\
     \cos 2\theta&=\frac{2\lambda_\phi v_\phi^2-2\lambda_Hv_H^2}{M_h^2-M_S^2},
\end{align}
such that ${\rm diag}(M_h^2,M_S^2)= R(\theta)^T M^2 R(\theta)$.  The scalar  masses are
\begin{align}
    M_{h,S}^2&=\frac{1}{2}\left[2\lambda_Hv_H^2+2\lambda_\phi v_\phi^2\mp\sqrt{\left[2\lambda_Hv_H^2-2\lambda_\phi v_\phi^2\right]^2+4\lambda_{SH}^2v_\phi^2v_H^2}\right],\\
    M_h^2&=2(v_H^2\lambda_H\cos^2\theta - \lambda_{SH}v_\phi v_H\cos\theta\sin\theta + \lambda_\phi v_\phi^2\sin^2\theta),\\
    M_{S}^2&=2(v_H^2\lambda_H\sin^2\theta + \lambda_{SH}v_\phi v_H\cos\theta\sin\theta+\lambda_\phi v_\phi^2\cos^2\theta)
\end{align}
where we assumed, without loss of generality, that $M_{h}<M_{S}$. $h$ can be identified with the Higgs boson observed at the LHC with a mass of about $125$ GeV \cite{ATLAS:2015yey}. 

From the minima conditions we can obtain $\mu_H$ and $\mu_\phi$, which are replaced back in $\mathcal{V}$:
\begin{align}
    \frac{\partial \mathcal{V}}{\partial \phi_R}=0&\to v_\phi\left(-\mu_\phi^2+\lambda_\phi v_\phi^2+\frac{1}{2}\lambda_{SH}v_H^2\right)=0,\\
    \frac{\partial \mathcal{V}}{\partial H^0_R}=0&\to v_H\left(-\mu_H+\lambda_H^2v_H^2+\frac{1}{2}\lambda_{SH}v_\phi^2\right)=0,
\end{align}
leading to
\begin{align}\label{eq:scalarnonfree}
    \lambda_H&=(M_{h}^2\cos^2\theta+M_{S}^2\sin^2\theta)/(2v_H^2),\\
    v_\phi&=(M_{S}^2-M_{h}^2)\sin\theta\cos\theta/(\lambda_{SH}v_H),\label{eq:vphi}\\
    \lambda_\phi&=\left(M_{h}^2\sin^2\theta+M_{S}^2\cos^2\theta\right)/(2v_\phi^2).\label{eq:laphi}
\end{align}
In this way, the parameters of the scalar potential can be written in terms of the physical parameters.

We can now write the fermion interaction term in the mass eigenstate basis,
\begin{align}\label{eq:L2}
 \mathcal{L}_\psi&=\,\,-y_s\overline{\psi}_L\psi_R\, \phi+ \rm{h.c.}\\
 &=\,\,-m_\psi\overline{\psi}\psi\left[1+(-h\,\sin\theta+ S\,\cos\theta)/v_\phi\right], 
 \end{align}
where 
\begin{align}\label{eq:ys}
    y_s=m_\psi/v_\phi. 
\end{align}
Thus, the chiral fields $\psi_L$ and $\psi_R$ form a massive Dirac fermion, $\psi=\psi_L+\psi_R$. Notice that $\psi$ is automatically stable and thus the dark matter particle in this model. It couples only to $h$ and $S$ and with an strength proportional to its mass. In addition, since $\theta$ must be  small (see section \ref{sec:pspace}),  $\psi$ couples more strongly to $S$ than to $h$.

Altogether, this model includes only 4 free parameters, which can be taken to be
\begin{align}
m_\psi, M_S,\sin\theta, \lambda_{SH}.
\end{align}
All other parameters are either known or can be determined from these using the equations above. Let us emphasize that no extra assumptions have been made to simplify the Lagrangian or the parameter space --this is the most general setup consistent with the particle content and the symmetries imposed. Given this restricted set of parameters, one may wonder whether it is actually possible to explain the observed density of dark matter while remaining consistent with known experimental and theoretical bounds. That is the question we will address in the next two sections.

\section{Dark matter phenomenology}
\label{sec:pheno}
\begin{figure}[t]
    \centering
    \includegraphics{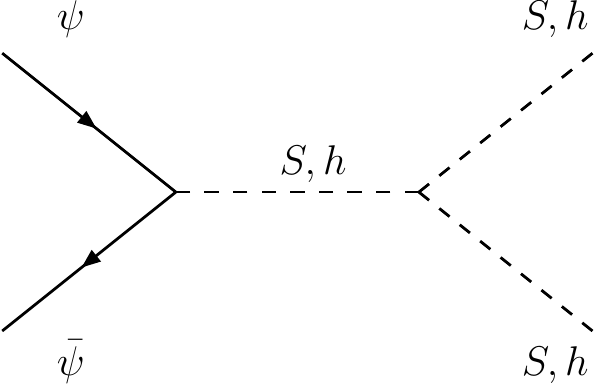}\hspace{1cm}
    \includegraphics{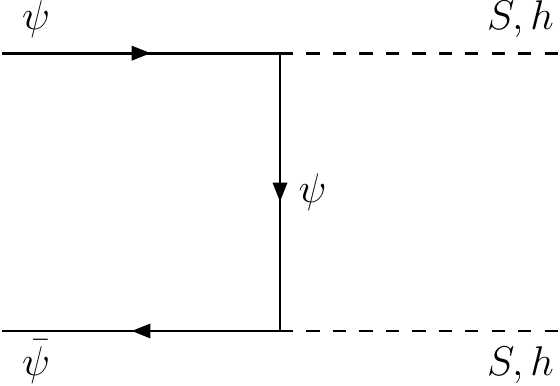}\\
    \caption{Some of the Feynman diagrams that contribute to dark matter annihilation in this model. In the left diagram, the final state scalars can be replaced by $W,Z$ or SM fermions.}
    \label{fig:annidiags}
\end{figure}

In this model, the dark matter particle couples directly only to $h$ and $S$, and through them, via mixing, to the other SM particles. 
For illustration, figure \ref{fig:annidiags} displays the Feynman diagrams that contribute to dark matter annihilation into two scalars. Throughout this work, the dark matter observables are calculated  by {\tt micrOMEGAs} version 5.0 \cite{Belanger:2018ccd} via {\tt LanHEP}~\cite{Semenov:2014rea}. Let us stress that {\tt micrOMEGAs} numerically solves the Boltzmann equation for $\psi$, automatically taking into account all the different processes that contribute to the  dark matter relic density, including decays and $2 \to 2$ scatterings.

\begin{figure}[t]
    \centering
    \includegraphics[scale=0.7]{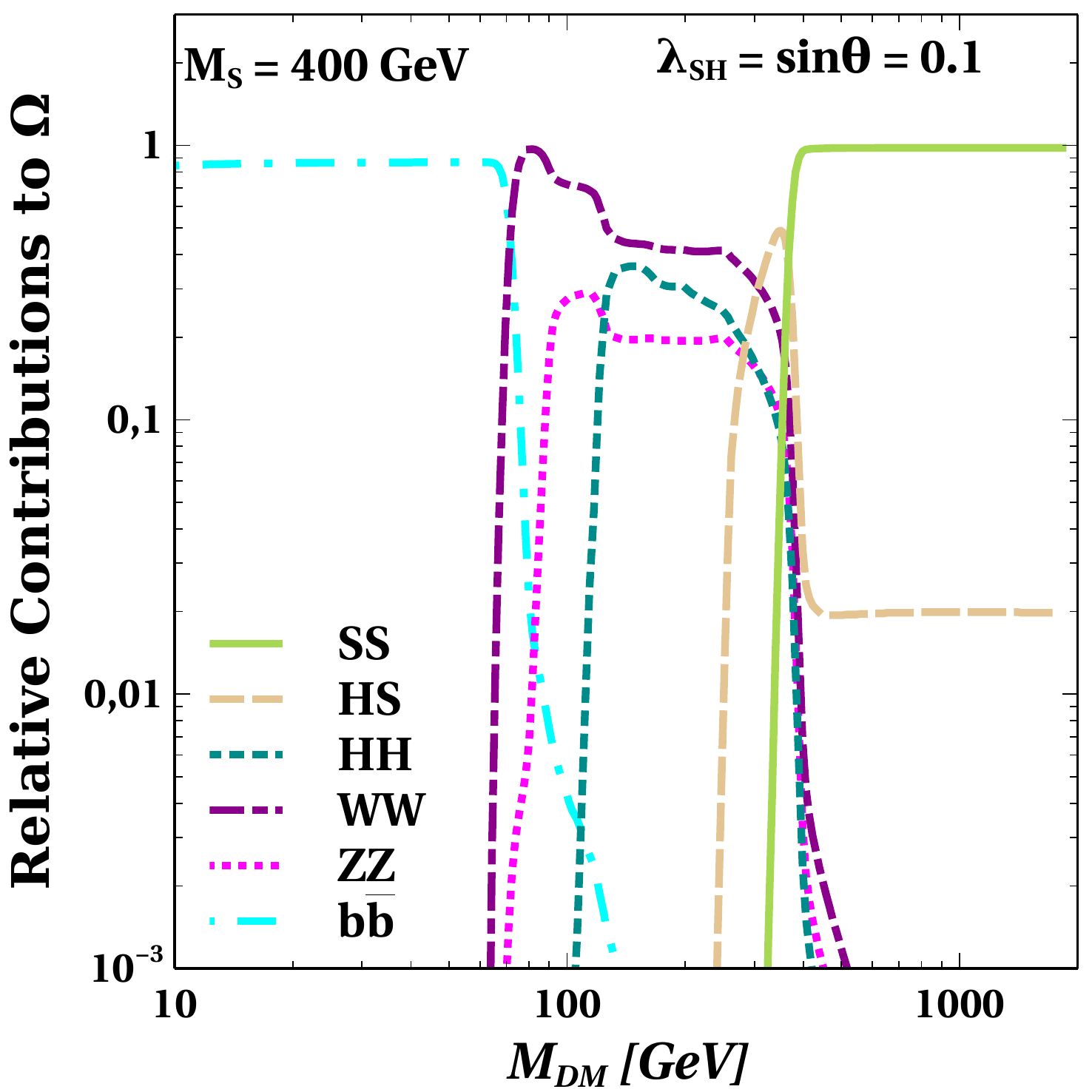}
    \caption{The relative contributions of different final states to the total relic density.}
    \label{fig:branchings}
\end{figure}

Figure \ref{fig:branchings} displays, as a function of the dark matter mass, the relative contributions of different final states to the relic density.  As anticipated, the dominant annihilation final state strongly depends on $M_\psi$, being $b\bar b$ for dark matter masses smaller than $M_h/2$ --below the Higgs resonance. Right above the Higgs resonance, other final states become relevant including  $W^+W^-$, $HH$ and $HS$. Finally, for $M_\psi\gtrsim M_S$ it is the $SS$ final state that entirely dominates the relic density. Even if this figure was obtained for a specific set of parameters --$M_S=400$ GeV, $\lambda_{SH}=\sin\theta=0.1$-- the pattern it reveals is quite generic. Notice, from the right diagram in figure \ref{fig:annidiags}, that the cross section for the annihilation into $SS$  goes like $y_s^4$ and since $M_\psi\propto y_s$, that process is expected to dominate at large dark matter masses.  In fact, as we will see in the next section, for \emph{all} the viable models that lie outside the resonances, the process that sets the dark matter relic density is  indeed $\psi\bar\psi\to SS$.

\begin{figure}[t]
    \centering
    \includegraphics[scale=0.7]{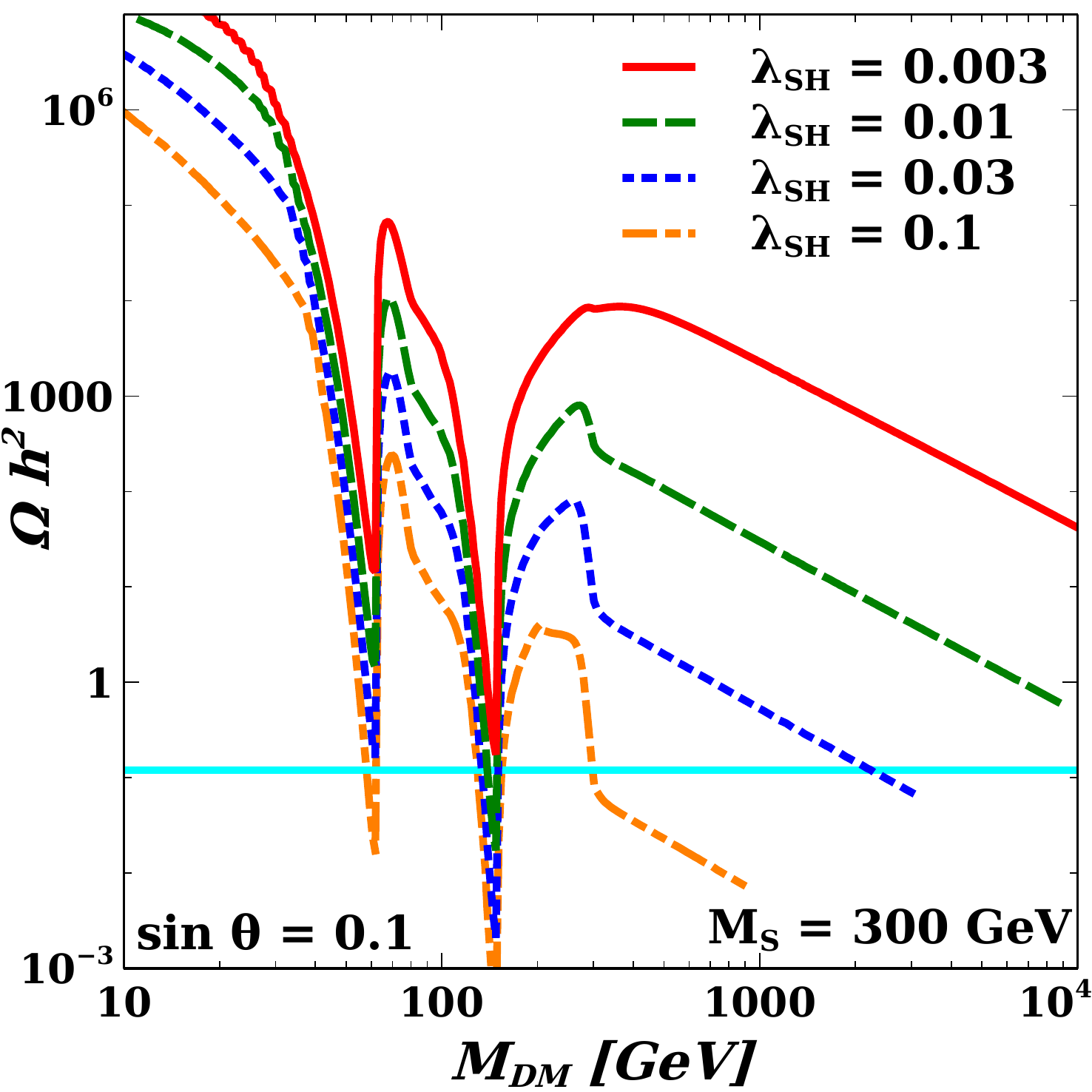}
    \caption{The relic density as a function of the dark matter mass for different values of $\lambda_{SH}$. From top to bottom the lines correspond to $\lambda_{SH}=0.003$ (solid, red), $0.01$ (dashed, green), $0.03$ (dotted, blue), and $0.1$ (dash-dotted, orange). For this figure $M_S=300$ GeV and $\sin\theta=0.1$. The (cyan) horizontal band corresponds to the observed dark matter density. }
    \label{fig:reliclash}
\end{figure}

Let us now study how the relic density varies with the free parameters of the model. Figure \ref{fig:reliclash} shows $\Omega h^2$ versus the dark matter mass for different values of the Higgs-portal coupling, $\lambda_{SH}$. The other two parameters were taken as $\sin\theta=0.1$ and $M_S=300$ GeV. Notice that $\lambda_{SH}$ has a direct impact on the dark matter interactions because, according to equations (\ref{eq:vphi}) and (\ref{eq:ys}), $y_s\propto \lambda_{SH}$. Thus, we expect $\Omega \propto \lambda_{SH}^{-2}$ at low dark matter masses whereas  $\Omega\propto \lambda_{SH}^{-4}$ at high dark matter masses, in agreement with what is observed in the figure. Two additional features are clearly observed in this plot. First, the suppression of the relic density at the Higgs and $S$ resonances --$M_\psi=62.5, 150$ GeV respectively. Second, the drop in the relic density once the $SS$ annihilation channel becomes kinematically accessible ($M_\psi\sim 300$ GeV). Finally, notice that agreement with the observed dark matter density~\cite{Planck:2018vyg} (horizontal cyan band) can be obtained, for this set of parameters, only if $\lambda_{SH}$ is not too small and if the dark matter mass lies either at the resonances or above $M_S$.

\begin{figure}[t]
    \centering
    \includegraphics[scale=0.7]{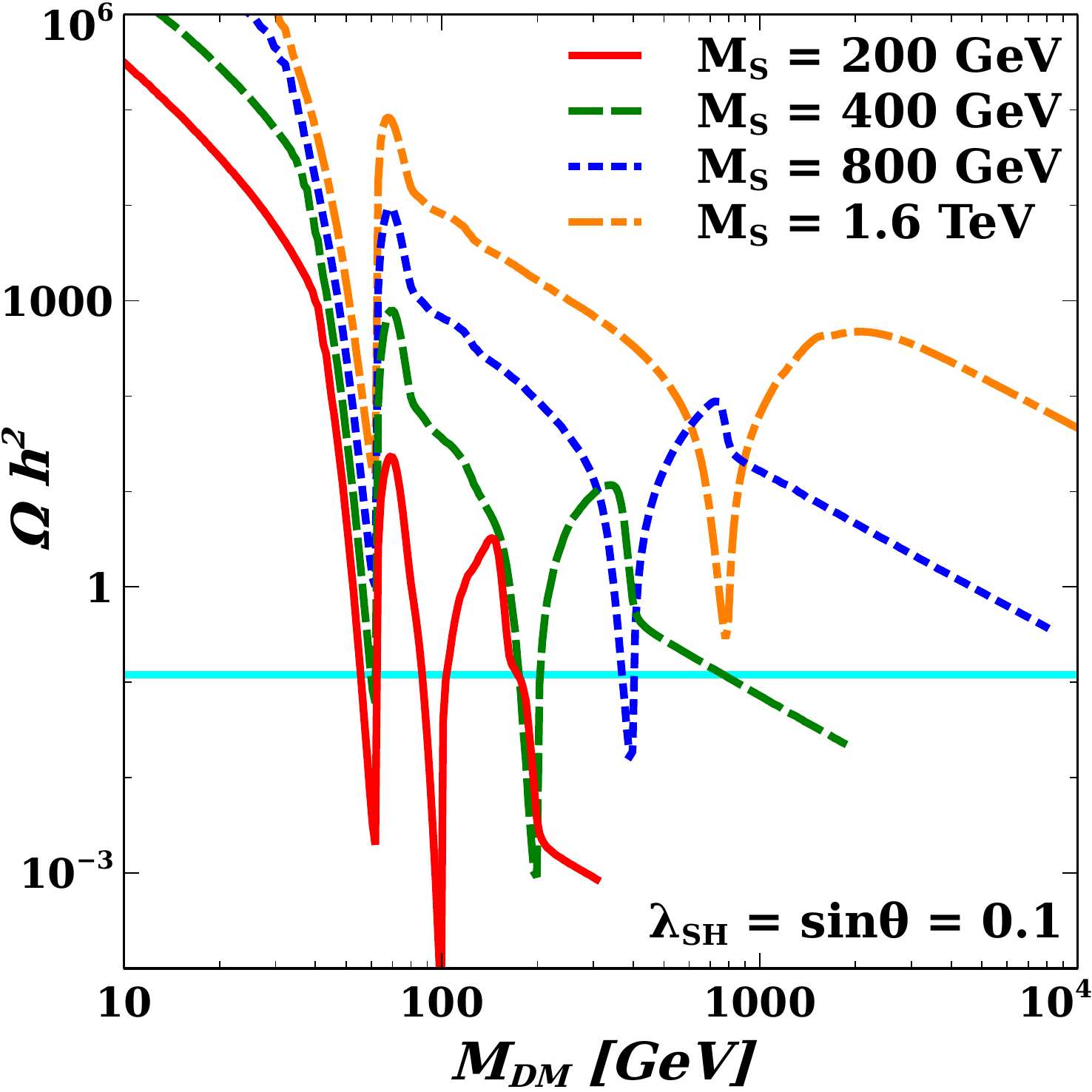}
    \caption{The relic density as a function of the dark matter mass for different values of $M_{S}$. From bottom to top the lines correspond to $M_{S}=200$ GeV (solid, red), $400$ GeV (dashed, green), $800$ GeV (dotted, blue), and $1.6$ TeV (dash-dotted, orange). For this figure  $\lambda_{SH}=\sin\theta=0.1$. The (cyan) horizontal band corresponds to the observed dark matter density.}
    \label{fig:relicms}
\end{figure}

\begin{figure}[t]
    \centering
    \includegraphics[scale=0.7]{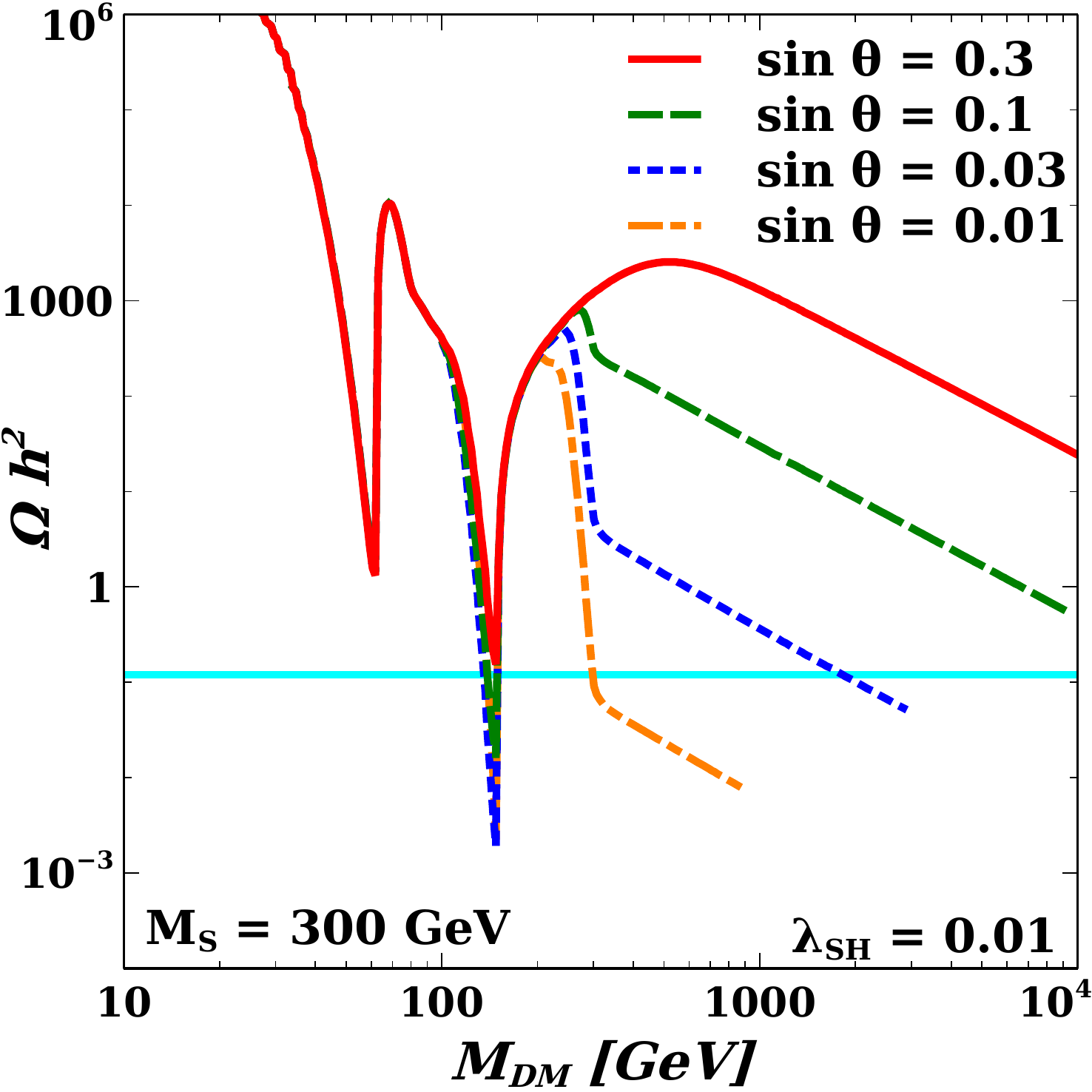}
    \caption{The relic density as a function of the dark matter mass for different values of $\sin\theta$. From top to bottom the lines correspond to $\sin\theta=0.3$ (solid, red), $0.1$ (dashed, green), $0.03$ (dotted, blue), and $0.01$ (dash-dotted, orange). For this figure $M_S=300$ GeV and $\lambda_{SH}=0.01$. The (cyan) horizontal band corresponds to the observed dark matter density.}
    \label{fig:relicst}
\end{figure}

Figure \ref{fig:relicms} shows instead the dark matter density for different values of $M_S$.  For this plot the dimensionless parameters $\lambda_{SH}$ and $\sin\theta$ were both set to $0.1$.  In this case, the position of the $S$ resonance moves from line to line, and the relic density tends to increases with $M_S$. Notice that for $M_S=1.6$ TeV the predicted relic density is always larger than the observed value (cyan horizontal band), whereas for the other cases it is not so and it becomes possible to satisfy the relic density constraint for some values of $M_{DM}$.  Finally, the variation of the relic density with $\sin\theta$ is illustrated in figure \ref{fig:relicst} for $M_S=300$ GeV and $\lambda_{SH}=0.01$. $\sin\theta$ affects the dark matter interactions mainly through $v_\phi$ --see equation (\ref{eq:vphi})--, which in turn determines $y_s$. At low dark matter masses, the dependence on $\sin\theta$ cancels out among the two vertices but at high masses we expect $\Omega h^2\propto \sin^4\theta$, in agreement with the plot. Note that the observed value of the dark matter density tends to prefer smaller values of $\sin\theta$.

\begin{figure}[t]
    \centering
    \includegraphics{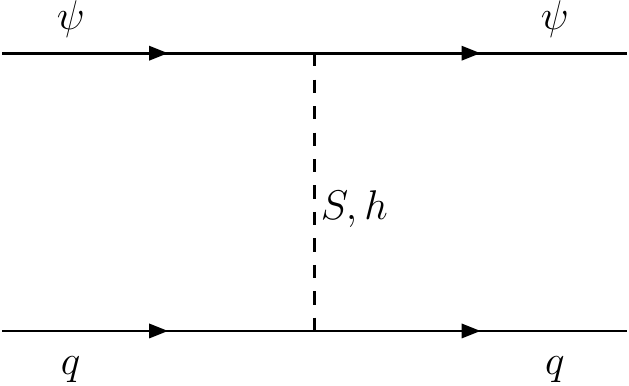}
    \caption{The Feynman diagrams that contribute to direct detection in this model.  }
    \label{fig:dddiags}
\end{figure}

Besides the relic density, another crucial observable  in this model is the dark matter scattering  cross section, which is subject to strong limits from direct detection experiments such as XENON-nT~\cite{XENON:2022ltv} and LZ~\cite{LZ:2022lsv}. Dark matter-quark scattering proceeds via a t-channel diagram mediated by $h$ and $S$ --see Figure \ref{fig:dddiags}. The resulting dark matter-nucleon cross section is spin-independent and given by  
\begin{equation}
\sigma_{SI}=\frac{\mu^2}{\pi}\left(\frac{y \sin\theta\cos\theta}{v_{H}}\right)^2\left(\frac{1}{M_h^2}-\frac{1}{m_S^2}\right)^2m_\mathcal{N}^2f_\mathcal{N}^2
\end{equation}
with $m_{\mathcal{N}}=939$ MeV, $f_{\mathcal{N}}\approx 0.3$ and $\mu=m_\psi m_\mathcal{N}/(m_\psi+m_\mathcal{N})$. From this information alone, however, it is not possible to say much about the importance of current direct detection bounds. To that end, we must first find the models that are consistent with the observed value of the dark matter density. That is precisely what we are going to do in the next section.

\section{The viable parameter space}
\label{sec:pspace}
In this section we present the main results of the exploration of the parameter space: the regions compatible with all the constraints are defined and characterized,  and the direct and indirect detection prospects are also analyzed. 

\subsection{Sampling}

\begin{table}[t]
    \centering
    \begin{tabular}{|c|c|c|}
        \hline  
         Free parameter & Range & Prior\\
        \hline
        $M_\psi$ & $[10, 10^4]$ GeV & Log\\ 
        \hline
        $M_S$ &  $[150, 10^4]$ GeV & Log\\ 
        \hline
        $\lambda_{SH}$ & $[10^{-4}, 10]$ & Log\\
        \hline
        $\sin\theta$ & $[10^{-5}, 1]$ & Log\\
        \hline
    \end{tabular}
    \caption{Intervals for the model's free parameters used with the differential evolution scan, alongside their corresponding prior likelihoods.}
    \label{tab:priors}
\end{table}

Our numerical analysis aims to make a statistically sound inference on the viable parameter space~\cite{AbdusSalam:2020rdj}. For this purpose, we adopt a frequentist approach and perform 4-dimensional scans using the differential evolution algorithm \cite{Storn1997,DBLP:journals/corr/abs-1907-10121}. The relevant outputs for a frequentist analysis, such as best-fit points and 68.3\% (1$\sigma$) and 95.4\% (2$\sigma$) confidence level (C.L.) regions, are created by maximizing the likelihood via Wilks’ theorem~\cite{Wilks:1938dza}. 
We assume the following likelihood function
\begin{align}
    \mathcal{L}(\vec{\alpha})=\mathcal{L}_{\Omega}(\vec{\alpha})\times\mathcal{L}_\text{DD}(\vec{\alpha}),
\end{align}
that accounts for the relic density measurement and the direct detection upper bound as a function of the free parameters $\vec{\alpha}=(M_\psi,M_S,\lambda_{SH},\sin\theta)$ --see Table \ref{tab:priors}. Note that for simplicity we take $M_S\geq 150$ GeV in our analysis. Moreover, we rely on the profile likelihood ratio (PLR), which is defined as
\begin{align}
    \Lambda=\frac{\mathcal{L}(\alpha_i)}{\mathcal{L}_{\rm max}},
\end{align}
where 
\begin{align}
    &\mathcal{L}(\alpha_i)=\underset{\alpha_i\notin \vec{\alpha}}{{\rm  max}}\,[\mathcal{L(\vec{\alpha})}],\hspace{1cm}\mathcal{L}_{\rm max}={\rm max}\,[\mathcal{L(\vec{\alpha})}],
\end{align}

We assume Gaussian likelihoods in such a way that their corresponding expressions read
\begin{align}
    \chi_\mathcal{O}^2(\vec{\alpha})=-2\ln \mathcal{L}_\mathcal{O}(\vec{\alpha})=\left(\frac{\mathcal{O}^{\text{th}}-\mathcal{O}^{\text{obs}}}{\Sigma_\mathcal{O}}\right)^2, 
\end{align}
where $\mathcal{O}^{\text{th}}$ and $\mathcal{O}^{\text{obs}}$ are the theoretical prediction and experimental measurement of the observable $\mathcal{O}$, respectively, and $\Sigma_\mathcal{O}$ is the associated total standard deviation.
From the dark matter abundance measurement reported by PLANCK~\cite{Planck:2018vyg}, 
\begin{align}
    \Omega_{\text{DM}}h^2=0.1198\pm 0.0012,
\end{align}
we associate the central value to $\Omega^{\text{obs}}=0.1198$. Moreover, we consider a model to be compatible with this measurement if the relic density obtained from {\tt micrOMEGAs}, $\Omega^{\text{th}}$, lies roughly between 0.11 and 0.13, that is,  within an estimated theoretical and experimental uncertainty of 10\%, $\Sigma_\Omega=0.1\Omega^{\text{obs}}$. 
Instead, to account for the current upper bounds reported by the LZ collaboration~\cite{LZ:2022lsv} a null mean value is adopted, $\sigma_\text{SI}^\text{obs}=0$, whereas $\sigma_\text{SI}^\text{th}$ is calculated through {\tt micrOMEGAS}. Finally,  $\Sigma^2_{\sigma_\text{SI}}=[0.2\sigma_\text{SI}^\text{th}]^2 + [\Sigma_{\sigma_\text{SI}}^\text{LZ}/1.64]^2$, where $0.2\sigma_\text{SI}^\text{th}$ corresponds to an estimated theoretical uncertainty of 20\% and $\Sigma_{\sigma_\text{SI}}^\text{LZ}$ is the 90\% confidence level upper bound (within $2\sigma$ sensitivity band) on the spin-independent cross section reported by LZ collaboration.  

We impose the following theoretical  constraints on the scalar potential parameters~\cite{Belanger:2014bga}:
\begin{itemize}
    \item Perturbatibity:
\begin{align}
\label{eq:pert}
    &|\lambda_H|<\frac{2}{3}\pi,\,\,\,|\lambda_\phi|<\pi,\,\,\,|\lambda_{SH}|<4\pi, 
\end{align}
\item Perturbative unitarity:
\begin{align}
   & |\lambda_H|<8\pi,\,\,\,|\lambda_\phi|<8\pi,\,\,\,
  \left|6\lambda_H+\lambda_{\phi}\pm \sqrt{(6\lambda_{H}-\lambda_\phi)^2+8\lambda^2_{SH}}\right|<16\pi.
\end{align}
\item Vacuum stability:
For this case the necessary and sufficient vacuum stability conditions are
\begin{align}
&\lambda_{H}>0,\,\,\,\lambda_\phi>0, \,\,\,\lambda_{SH}+2\sqrt{\lambda_H\lambda_\phi}>0.
\end{align}
\item Global vacuum $(v_H,v_S)$:
\begin{align}
    &\lambda_{SH}v_\phi^2 v_H^2+\lambda_\phi v_\phi^4 >0,\,\,\,\,\,\lambda_H v_H^4+\lambda_{SH}v_\phi^2 v_H^2 > 0.
\end{align}
\end{itemize}
Besides, the Yukawa coupling is subject to the perturbativity constraint 
\begin{align}\label{eq:yuk-pert}
    y_s<\sqrt{4\pi}.
\end{align} 

Concerning collider constraints, since the Higgs signal strength gets modified by a factor of $\cos^2\theta$, it follows that $|\sin\theta|\lesssim 0.3$ for $M_{S}> M_h$  \cite{Falkowski:2015iwa,Arcadi:2019lka,Ferber:2023iso} from the most recent LHC results \cite{ATLAS:2022vkf,CMS:2022dwd}. Direct searches for additional Higgs bosons give stronger bounds on $\sin\theta$ than the Higgs signal strength for $150\,{\rm GeV}< M_{S} \lesssim 700$ GeV~\cite{Ferber:2023iso}. Models that do not satisfy these collider bounds or the theoretical constraints mentioned above are excluded from our sample, independently of their likelihood.

Our subsequent analysis is based on a set of  $\mathcal{O}(10^5)$ points,  which were obtained using the likelihood-driven sampling technique described before.  From this sample of points, the one-dimensional PLR can be computed for the most  relevant variables  --see e.g. figure \ref{fig:1DPLR}--, and the $2\sigma$ C.L. isolikelihood regions can be obtained via the projection onto several 2-dimensional planes --e.g. in figures \ref{fig:freepar}-\ref{fig:cxSI}--, as shown next.  Note that, since no sample is complete, there could be some models that  do not appear in our results, which may result in slight changes to the shown regions. The C.L. contours (the boundary of the regions formed by the selected points) reflect the models’s capability to explain the dark matter observables.  We construct $2\sigma$ C.L. regions projected onto several planes (binned into $150\times 150$ cells) by imposing the criteria $-2\ln\left(\mathcal{L}(M_\psi,M_S)/\mathcal{L}_{{\rm max}}\right)\leq 4\,(6.18)$ for one (two) free parameter(s) involved and profiling over the remaining parameters. 

\subsection{Results}
\begin{figure}[t]
    \centering
    \includegraphics[scale=0.4]{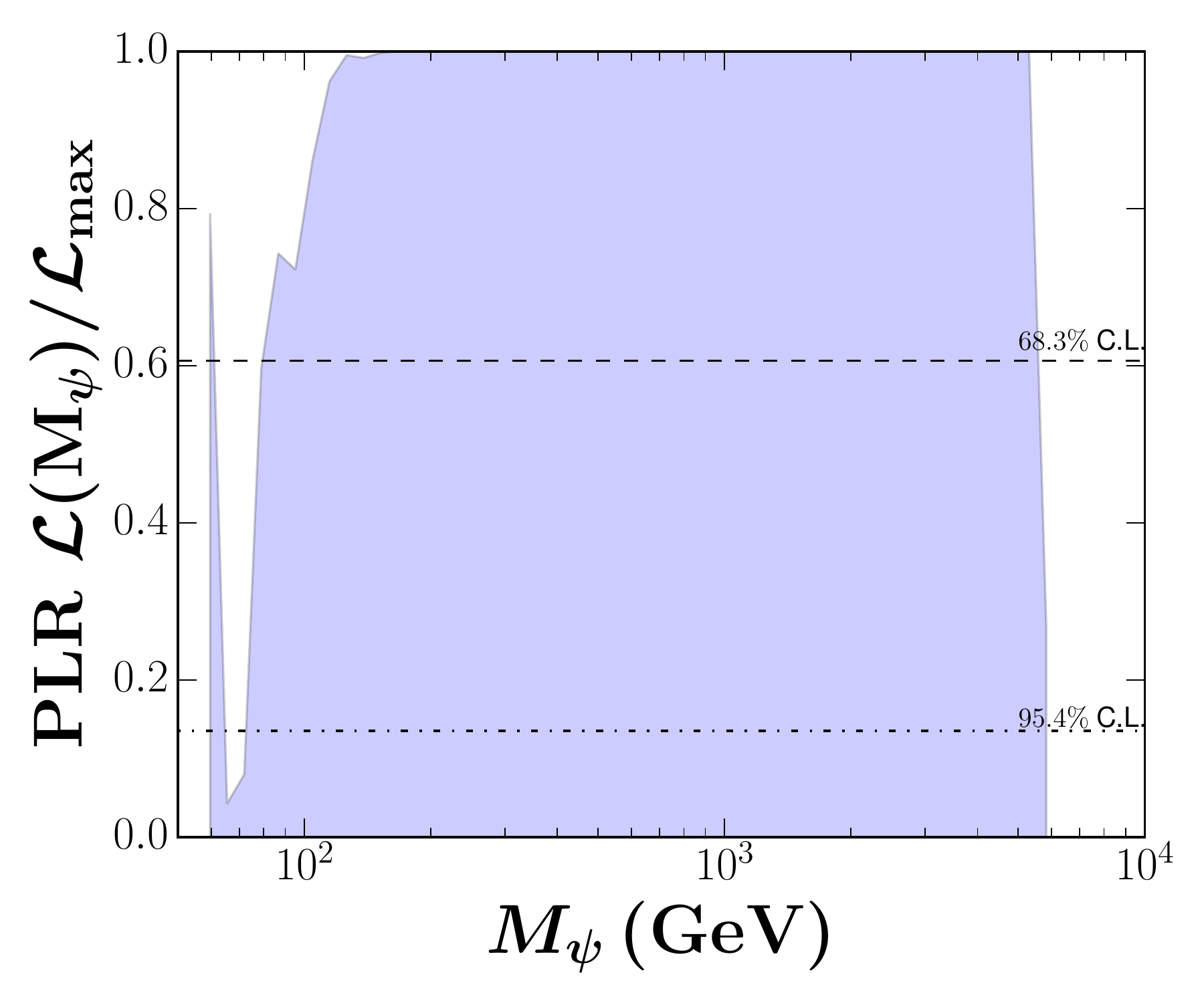}
    \includegraphics[scale=0.4]{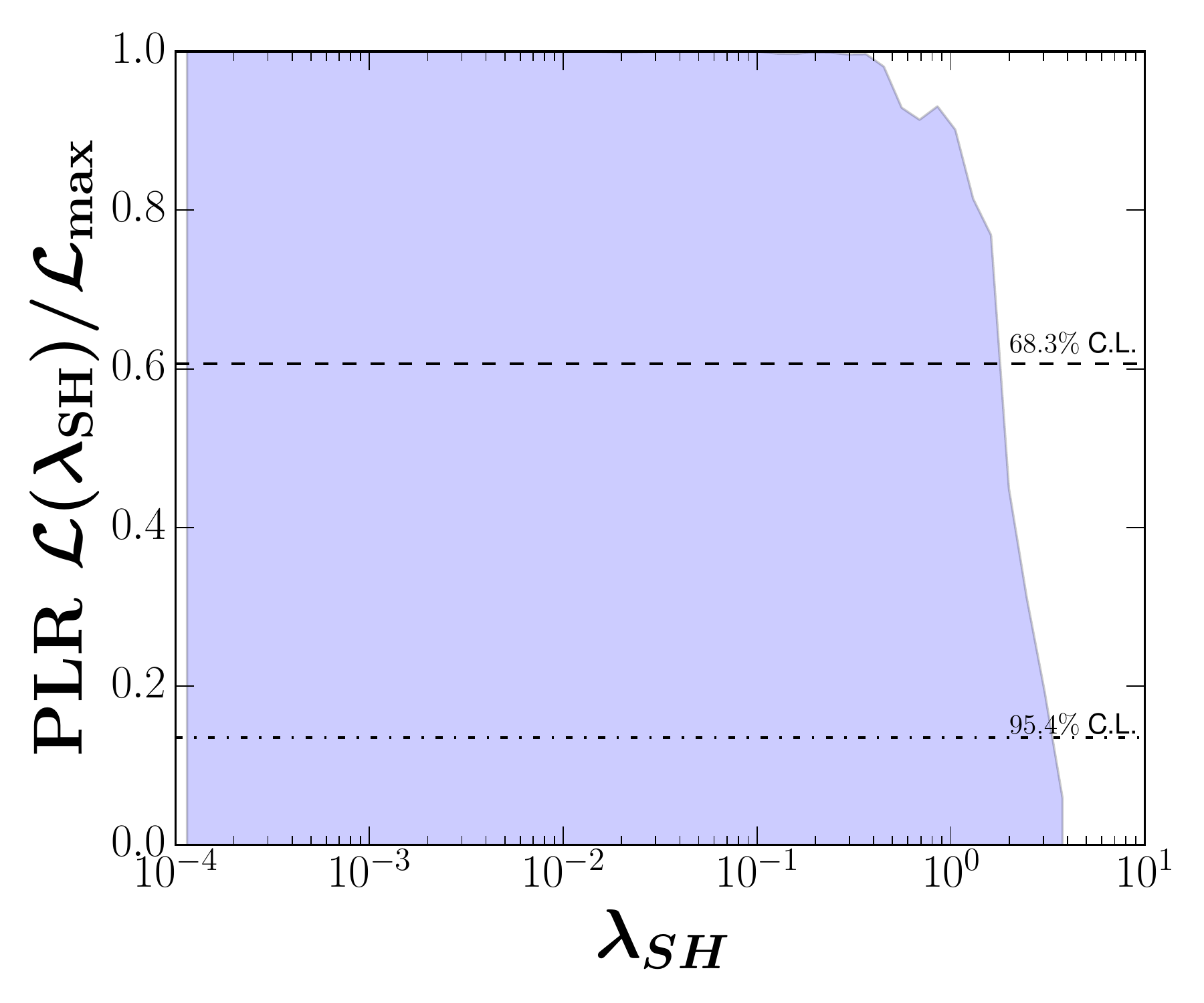}
    \caption{One-dimensional PLR for $M_\psi$ (left panel) and $\lambda_{SH}$ (right panel). Dashed and dot-dashed lines correspond to $1\sigma$ and $2\sigma$ confidence intervals, respectively. }
    \label{fig:1DPLR}
\end{figure}
\begin{figure}[t]
    \centering
    \includegraphics[scale=0.4]{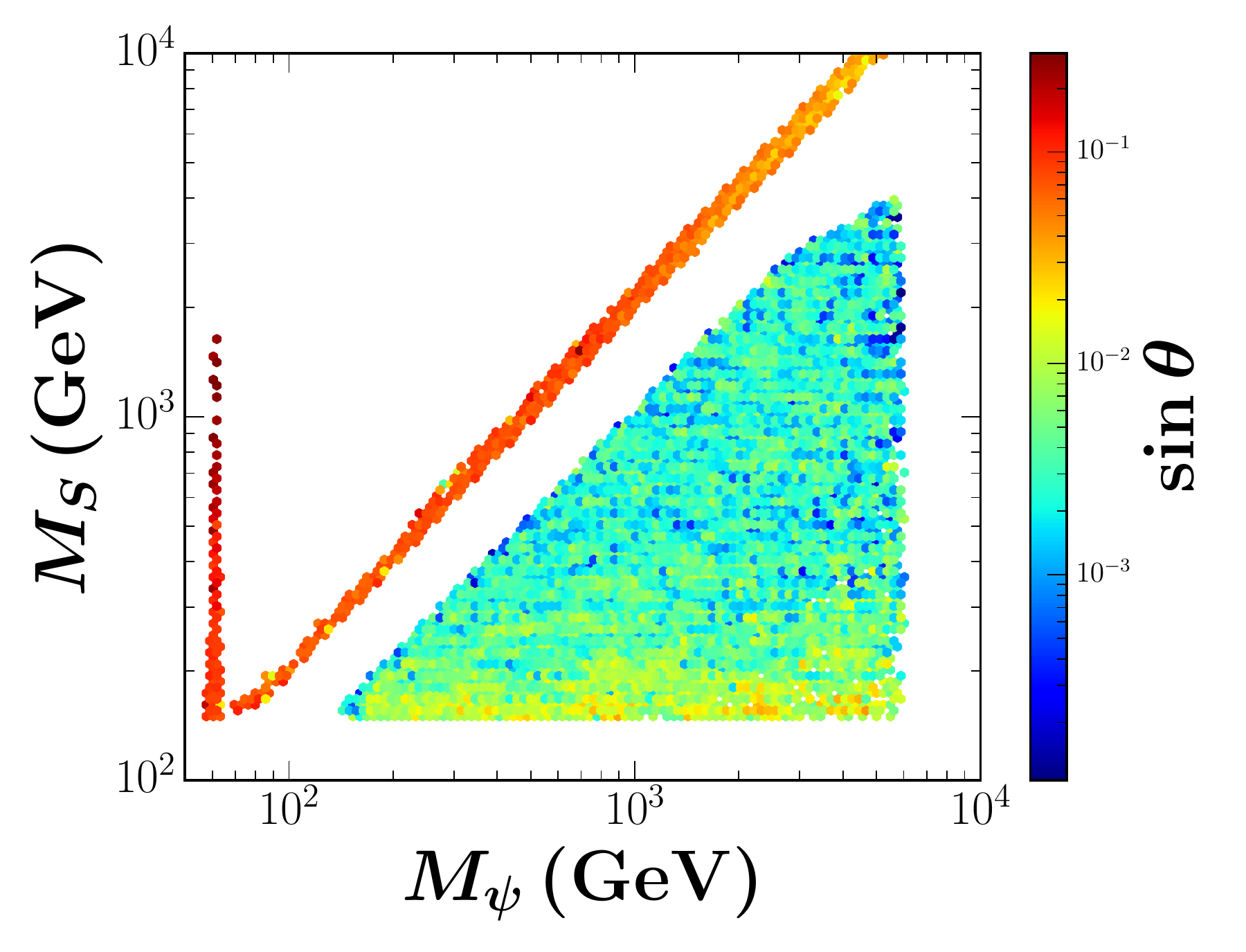}
    \includegraphics[scale=0.4]{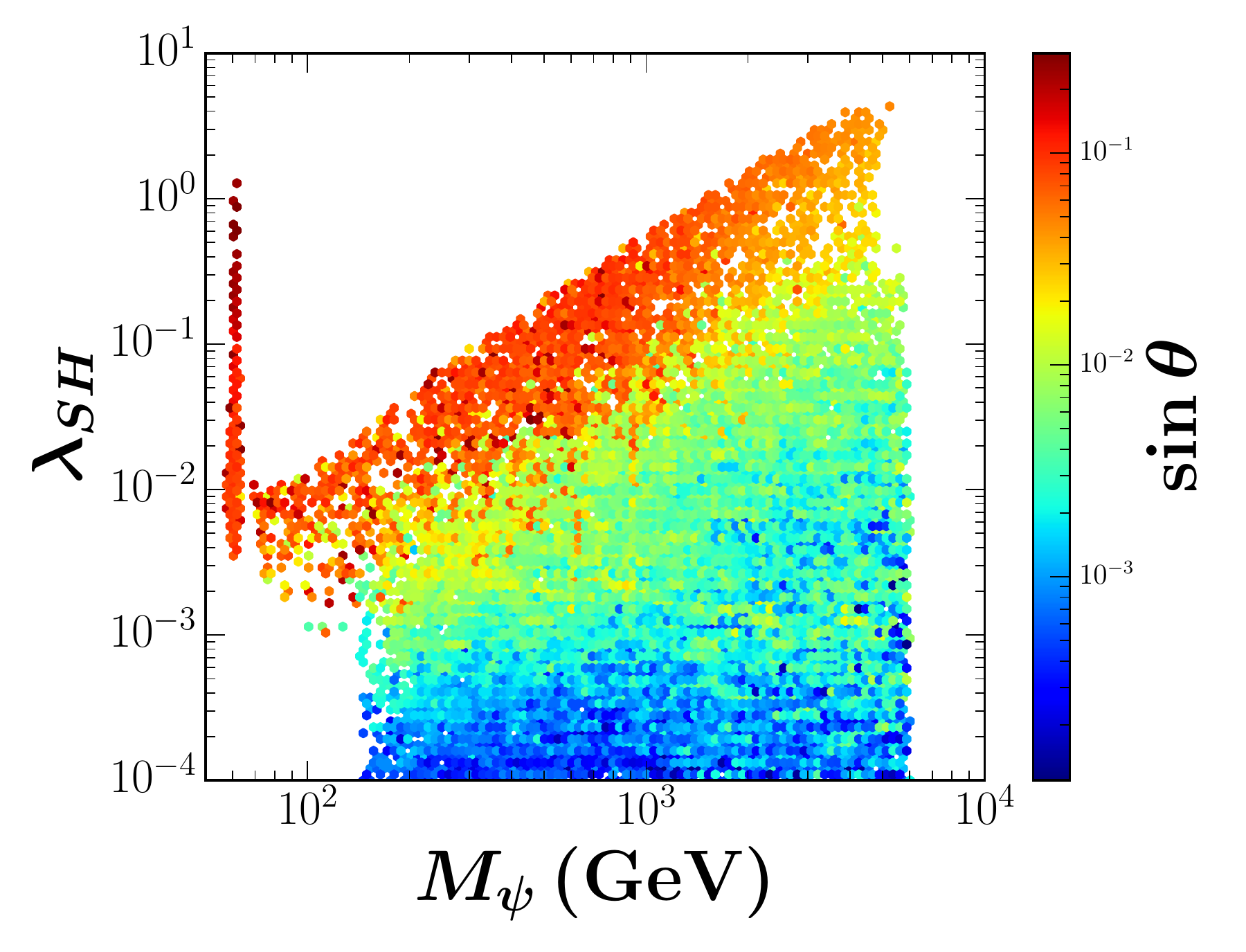} 
    \caption{$2\sigma$ C.L. regions in the planes of free parameters of the model $(M_\psi,M_S)$ and $(M_\psi,\lambda_{SH})$. 
    }
    \label{fig:freepar}
\end{figure}

The best-fit point of the sample (the set of parameter value having the maximum likelihood) is found to has the values $M_\psi=2000.18$ GeV, $M_S=877.15$ GeV, $\lambda_{SH}=3.33\times10^{-4}$ and $\sin\theta=1.02\times10^{-4}$. 
The one-dimensional PLR obtained from the 4-dimensional scans are displayed in Fig.~\ref{fig:1DPLR} for $M_\psi$ and $\lambda_{HS}$.  
These are obtained by calculating the $\mathcal{L}(M_\psi)/\mathcal{L}_{\rm max}$ and $\mathcal{L}(\lambda_{SH})/\mathcal{L}_{\rm max}$ for the whole sample of points.  Then the $1\sigma$ ($2\sigma$) C.L. region becomes the region bounded from below by $-2\ln\left(\mathcal{L}(\alpha_i)/\mathcal{L}_{{\rm max}}\right)\leq 1\,(4)$.  
From these results it follows that the global fit slightly prefers dark matter masses above the Higgs mass and Higgs portal couplings below $0.5$.  The other two 1D PLR, for $M_S$ and $\sin\theta$, are not shown since they are not constrained by the fit --the PLR is basically flat and equal to 1 within the whole allowed range. Similarly, the $1\sigma$ and $2\sigma$ C.L. regions in the 2-dimensional planes involving free parameters are mostly unconstrained by our global fit except in the white zones of Fig.~\ref{fig:freepar}.

  The allowed $2\sigma$ C.L. regions of this model are shown in figure \ref{fig:freepar}, projected onto the plane $(M_\psi,M_S)$ in the left panel, and $(M_\psi, \lambda_{SH})$ in the right panel, with the value of $\sin \theta$ color coded. Remarkably, they span a range of two orders of magnitude in the dark matter mass, from $60$ GeV to $6$ TeV approximately.  In the left panel, three areas  can be clearly differentiated, depending on how the relic density constraint is satisfied: the vertical line at $M_\psi\sim M_h/2$ is due to the Higgs resonance; the diagonal line extending  up to dark matter masses of $5$ TeV corresponds to the $S$ resonance; and the remaining triangular region comprises the models where the relic density is obtained via the annihilations $\psi\bar\psi\to SS$. Notice that this region prefers small values of $\sin\theta$, as explained in the previous section, whereas  the two resonances favor larger values. From the right panel, we also see that, as expected, $\lambda_{SH}$ tends to be larger at the resonances. In fact, outside the resonances we find $\lambda_{HS}\lesssim 0.5$, a fact that, as shown in the next subsection, has profound implications for the detection prospects in this model.  

\begin{figure}[t]
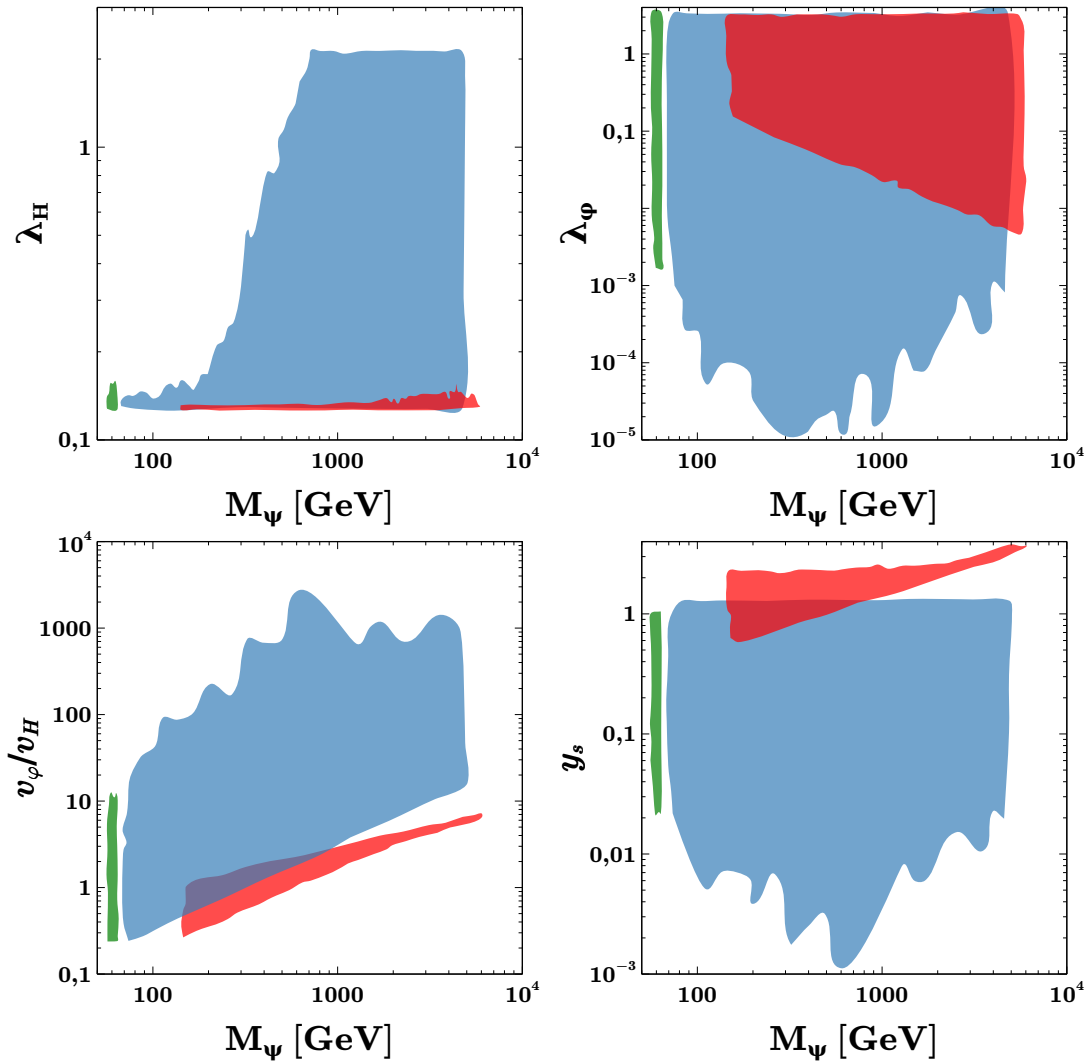

    \centering
    \includegraphics[scale=0.47]{Plots/fig9-1.pdf}
    \includegraphics[scale=0.47]{Plots/fig9-2.pdf}\\ 
    \includegraphics[scale=0.47]{Plots/fig9-4.pdf}
    \includegraphics[scale=0.47]{Plots/fig9-3.pdf}
    \caption{$2\sigma$ C.L. regions involving non-free parameters $(\lambda_H, \lambda_\phi, v_\phi, y_s)$ as a function of the dark matter mass according to their belongingness to the region I (red plus symbols), II (blue dots) or III (green crosses).  These regions are classified according to the mechanism responsible for achieving most of the relic density: Higgs resonance (region I), $S$ resonance (region II), and outside the resonances or equivalently $\psi\bar\psi\to SS$ (region III). }
    \label{fig:nonfreepar}
\end{figure}

It is also useful to know the allowed values for the parameters $(\lambda_H, \lambda_\phi, v_\phi, y_s)$, even if they are not free --Eqs.~(\ref{eq:scalarnonfree})-(\ref{eq:ys}).  In figure~\ref{fig:nonfreepar} we display the $2\sigma$ C.L. regions projected onto these parameters vs $M_\psi$ and at the same time classify the sample according to the mechanism responsible for their relic density: green for the Higgs resonance, blue for the $S$ resonance, and red for outside the resonances (equivalently $\psi\bar\psi\to SS$).  Notice from the top-left panel that for the resonant regions $\lambda_{H}$ may exceed significantly its SM value ($0.127$), providing an indirect way to probe this scenario. From the bottom-right panel, instead, we can see that $y_s$ must be large when the relic density is determined via $\psi\bar \psi\to SS$, as it is the coupling that fixes that annihilation cross section.

\subsection{Dark matter detection prospects}
\begin{figure}[t]
    \centering
    \includegraphics[scale=0.7]{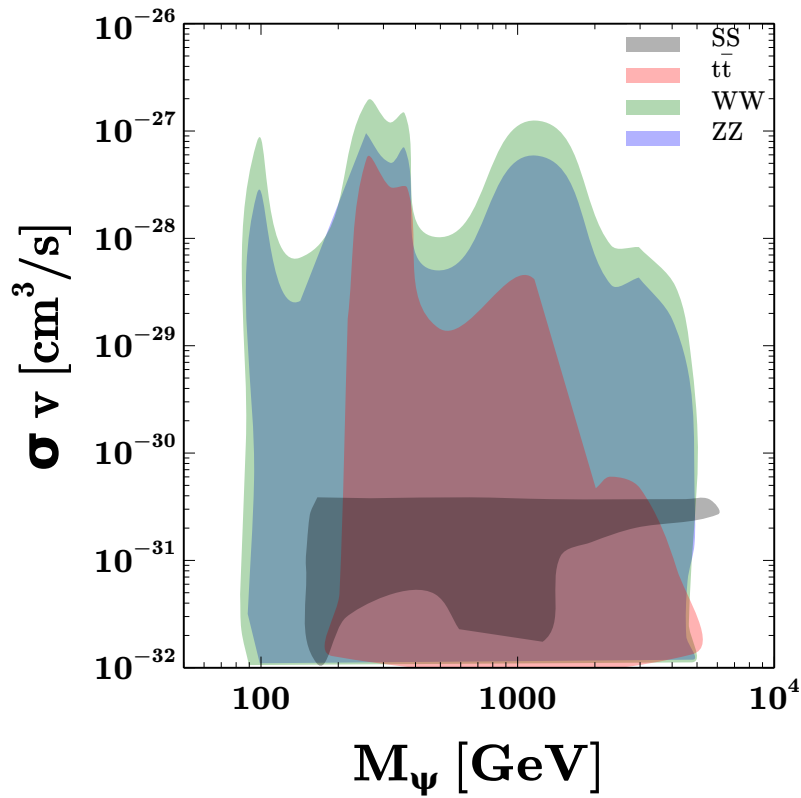}
    \caption{The dark matter annihilation rates as a function of $M_\psi$ for the most relevant final states within the $2\sigma$ C.L. region. The colored plus symbols correspond to viable models inside $S$-resonance while the black dots to those outside this.  In all cases, $\sigma v$ was computed for $v=1.3\times 10^{-3}$, which is typical for the Galactic halo. Due to the velocity suppression, the rates lie well below the thermal value of $3\times 10^{-26}$ cm$^{3}/$s.}
    \label{fig:sigmav}
\end{figure}

Now that the viable regions of this model have been determined, it is time to look at their implications, in particular regarding  dark matter detection searches.  Figure~\ref{fig:sigmav} displays the dark matter annihilation rate, $\sigma v$, for the most important final states. For this figure, $v\sim 10^{-3}$, which is typical for dark matter particles in the Galactic halo. Being $p$-wave suppressed, the annihilation into $SS$ features a cross section about five orders of magnitude below the thermal value, at around $3\times10^{-31}$cm$^3$/s. For models at the $S$ resonance, the cross section can be larger but they all lie well below the thermal value and below the sensitivity of current experiments \cite{Fermi-LAT:2015att,Reinert:2017aga,AMS:2016oqu}. Indirect detection searches, therefore, do not currently constrain this model and are not expected to do so in the near future.
 
\begin{figure}[t]
    \centering
    \includegraphics[scale=0.7]{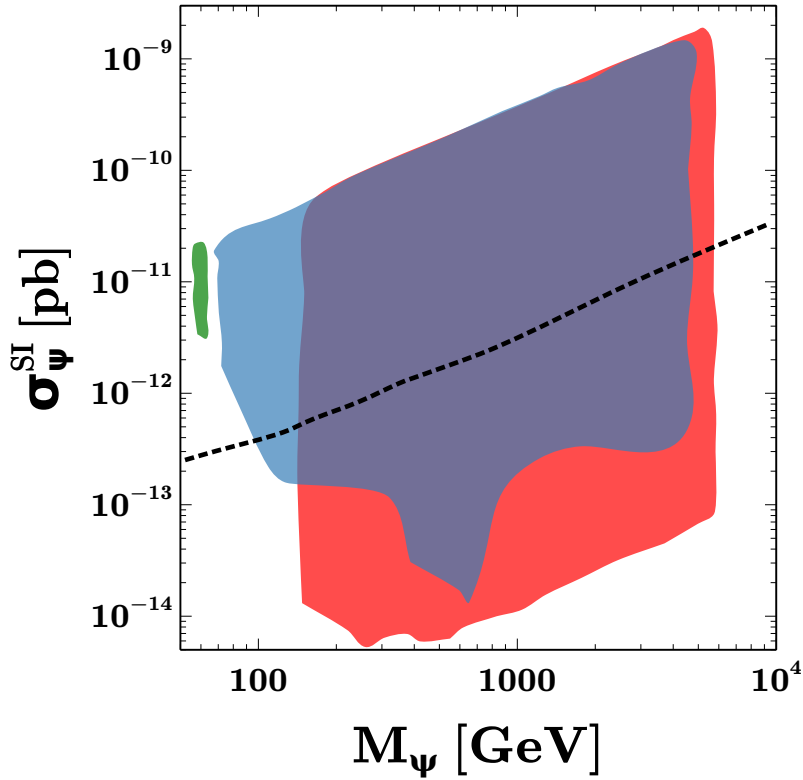}
    \caption{$2\sigma$ C.L. region for the spin-independent cross-section for elastic scattering of $\psi$ with nuclei as a function of $M_\psi$ in regions I, II and III. The dashed line shows the projected sensitivity of the DARWIN experiment~\cite{DARWIN:2016hyl}. }
    \label{fig:cxSI}
\end{figure}

More interesting is the case for direct detection. Figure~\ref{fig:cxSI} shows the viable regions projected onto the plane ($M_\psi$, $\sigma^{SI}$), with the color indicating the mechanism responsible for the relic density, as before. By construction, these regions are consistent with the most recent results from the  LZ experiment, which sets the upper bound on the spin-independent cross section.  The dotted line  corresponds to the expected sensitivity of the DARWIN experiment~\cite{DARWIN:2016hyl}.  Remarkably, this model can give rise to direct detection signals over the whole range of viable dark matter masses. In fact, the entire Higgs-resonance region will  be excluded if no events are observed at DARWIN. And  most of the $S$-resonance region will be probed by that experiment. The fact that the resonances are easier to test in direct detection experiments is related to the value of $\lambda_{HS}$, which is higher in such regions, as shown previously. Regarding the out of resonances region, we see from the figure that, even if not guaranteed,  a significant fraction of models may also induce an observable signal in DARWIN.  Hence, dark matter direct detection in this model looks very encouraging  and it   provides the most promising way to test it in the near future.  

\section{Discussion}
\label{sec:disc}

As we have seen, the $Z_6$ model we proposed is a simple, viable and testable scenario for Dirac dark matter. Let us briefly contrast this scenario against the $Z_2$-invariant singlet fermion model often considered in the literature --e.g.,~\cite{Kim:2008pp,Baek:2011aa,Baek:2012uj,Freitas:2015hsa,Lopez-Honorez:2012tov,Esch:2013rta}. In it, the fermion is odd while the scalar is even under the $Z_2$ so that the Lagrangian includes the following extra terms
\begin{align}\label{eq:VZ2}
    \mathcal{L}_{Z_2}\supset-m_D\bar{\psi}\psi -\left[ \mu^3_1 \phi  + \frac{1}{2}\mu_{2} \phi |H|^2 + \frac{1}{3}\mu_3 \phi^3 + \rm{h.c.}\right]. 
\end{align}
Even if the $\mu_1$ term can be set to zero under a shift of the singlet field, the extra scalar parameters $\mu_{2}$ and $\mu_3$ induce additional scalar interactions that give rise to fermion self-annihilations and elastic scatterings with nuclei. Moreover, a bare  mass term is now allowed for the dark matter ($m_D$ above), which means that $m_\psi$ and $y_s$ become independent of each other (see Eq.~(\ref{eq:yuk-pert})). Notice also that since bare Majorana mass terms  for $\psi_L$ and $\psi_R$ are allowed by the $Z_2$,  additional assumptions would be required to obtain Dirac dark matter. Clearly, the  $Z_6$ model studied here is simpler and more predictive.

One may wonder why do we impose a $Z_6$ and not some other $Z_N$ symmetry. It turns out that, for our particle content,  the $Z_6$ is the lowest $Z_N$ symmetry that can generate Dirac dark matter via spontaneous symmetry breaking --a $Z_2$ or a $Z_4$ would instead induce Majorana masses. Having three non-equivalent and non-trivial charges, each for the chiral fields $\psi_{L,R}$ and the scalar field $\phi$,  the $Z_6$ forbids bare mass terms for the fermions, either of  Dirac or of  Majorana type, while allowing a Yukawa interaction among all three fields. This interaction not only induces a Dirac mass after symmetry breaking but it also links the dark matter to the known particles, as shown before.  The $Z_6$ is thus quite special in this context.

An important feature of the $Z_N$ symmetries is that they may naturally appear as remnants of spontaneously broken $U(1)$ gauge symmetries. It is natural, therefore, to ask whether the $Z_6$ model investigated here can be incorporated into a $U(1)$ gauge extension of the SM. Besides a complex scalar (rather than real), such a scenario would demand an extended fermion sector to ensure  the cancellation of $U(1)$ anomalies. Due to the new gauge interactions and the extra fermion fields, which may constitute a second dark matter sector~\cite{Bernal:2018aon},  the phenomenology of this gauge model is expected to differ from the $Z_6$ scenario we discussed.  Another interesting aspect we have not addressed is the fact that the reduction in the number of free parameters (with respect to the $Z_2$-variant) may substantially modify those regions of the parameter space suitable for a strong first-order electroweak phase transition~\cite{Fairbairn:2013uta,Li:2014wia,Ghorbani:2017jls,Beniwal:2018hyi,Gould:2023jbz}. Hence,  the possibility of realizing electroweak baryogenesis~\cite{Ghorbani:2017jls,Beniwal:2018hyi} as well as the  signatures of this  model at current and future gravitational wave experiments~\cite{Beniwal:2018hyi} need to be studied in a future work.

\section{Conclusions}
\label{sec:conc}
One of the most pressing problems in fundamental physics today is determining the correct Standard Model  extension that accounts for the dark matter. In this work  we analyzed a new concise model in which the dark matter particle is a Dirac fermion, singlet under the SM gauge group.  This model includes only two new particles --the dark matter and a real scalar field-- and four free parameters, which renders it very predictive.  Two features set this scenario apart from related WIMP models. First, the dark matter stabilizing symmetry is a $Z_6$ rather than the standard $Z_2$. Second, the mass term for the dark matter particle is generated through the spontaneous breaking of the $Z_6$ symmetry. After studying its dark matter phenomenology, the viable parameter space of this model was explored using a profile likelihood approach. Our results indicate that this scenario remains viable over a wide range of  dark matter masses, extending well into the multi-TeV region.  Three mechanisms can successfully account for the observed dark matter abundance: annihilation at the Higgs resonance ($M_\psi\approx 62.5$ GeV), annihilation at the new scalar resonance ($M_\psi\approx M_S/2$), and annihilations into the new scalars, $\psi\bar\psi\to SS$, ($M_\psi>M_S$). All of them will be probed, either completely or partially,  by future direct detection experiments. Thus, this $Z_6$ scenario  not only offers one of the minimal dark matter extensions of the Standard Model, but also presents the most compelling  realization of singlet Dirac dark matter.

\section*{Acknowledgments}
OZ and CY received funding from the Patrimonio Autónomo - Fondo Nacional de Financiamiento para la Ciencia, la Tecnología y la Innovación Francisco José de Caldas (MinCiencias - Colombia) grant 82315-2021-1080.  OZ has been partially supported by Sostenibilidad-UdeA and the UdeA/CODI Grants 2020-33177, 2022-52380 and  2023-59130.

\bibliography{main}

\end{document}